# Enhanced Heat Flow between Charged Nanoparticle and Aqueous Electrolyte

Reza Rabani [a*], Mohammad Hassan Saidi [b], Ali Rajabpour [c], Laurent Joly [d], Samy Merabia [d‡]

[a] Department of Mechanical Engineering, Karaj Branch, Islamic Azad University, Karaj 3149968111, Iran
[b] Center of Excellence in Energy Conversion (CEEC), School of Mechanical Engineering, Sharif University of Technology, Tehran 11155-9567, Iran
[c] Advanced Simulation and Computing Laboratory (ASCL), Mechanical Engineering Department, Imam Khomeini International University, Qazvin 34148-96818, Iran
[d] Univ Lyon, Univ Claude Bernard Lyon 1, CNRS, Institut Lumière Matière, F-69622, Villeurbanne, France
Emails:  reza.rabani@iau.ac.ir (R.Rabani[*]),  samy.merabia@univ-lyon1.fr (S.Merabia[‡])

**ABSTRACT:** Heat transfer through the interface between a metallic nanoparticle and an electrolyte solution, has great importance in a number of applications, ranging from nanoparticle-based cancer treatments to nanofluids and solar energy conversion devices. However, the impact of the surface charge and the dissolved ions on heat transfer has been scarcely explored so far. In this study, we compute the interface thermal conductance between hydrophilic and hydrophobic charged gold nanoparticles immersed in an electrolyte using equilibrium molecular dynamics simulations. Compared with an uncharged nanoparticle, we report a threefold increase of the Kapitza conductance for a nanoparticle surface charge of +320 mC/m$^2$. This enhancement is shown to be approximately independent of surface wettability, charge spatial distribution, and salt concentration. This allows us to express the Kapitza conductance enhancement in terms of surface charge density on a master curve. Finally, we interpret the increase of the Kapitza conductance as a combined result of the shift of the water density distribution toward the charged nanoparticle and an accumulation of the counter-ions around the nanoparticle surface which increase the Coulombic interaction between the liquid and the charged nanoparticle. These considerations help to apprehend the role of ions in heat transfer close to electrified surfaces.

## INTRODUCTION

Investigating heat transfer in the vicinity of nanostructures immersed in water is of great importance for various applications[1–6]. These applications cover the fields of biomedicine[7–14], but also thermal managements[15–20]. Illuminated colloidal metal nanostructures indeed provide tools for local nanoscale heat generation, with developments in solar energy conversion[21–24], and also for thermally assisted destruction of biological cell[25,26]. These processes all rely on the unique property of metallic nanoparticles to strongly absorb light in the vicinity of their plasmon resonance frequency[27]. This extraordinary property opens the way for generating highly localized hot spots in a liquid environment. Thermal transport between an object and its environment is best quantified by the interfacial thermal conductance, also known as Kapitza conductance $G = q/\Delta T_{sl}$, with $\Delta T_{sl}$ the temperature jump at the interface and q the heat flux density. This conductance is indeed present at an interface between any pair of materials and is primarily related to the acoustic mismatch between the two media. In the case of colloidal nanoparticles, Kapitza conductance plays a key role in the heating kinetics of the surrounding medium[28–30].

Interfacial heat transfer at solid-liquid interfaces is governed by several parameters, including interfacial roughness[31,32], coating by polymer chains[33,34] and surfactants[35], and liquid pressure[36,37] among others. These parameters influence heat transfer not only at flat interfaces but also at spherical interfaces – including droplets and nanoparticles[38–43]. For nanoparticles, attention has been paid to the addition of polymers or surfactants and the influence of nanoparticle morphology on interfacial conductance[14,44]. To get a physical picture of interfacial heat transfer at liquid-solid interfaces, a few studies aimed at building connections between $G$ and the liquid structure close to the interface. For a water-graphene interface[45], $G$ is found to be inversely proportional to the height of the first density peak while for water inside charged boron nitride nanotube[46], the changes in $G$ are assigned to the strength of the hydrogen bonding, and not to water structuring. The relation between $G$ and the height of the first peak of water density is, however, found to be not universal[37,41].

A situation of high interest but poorly characterized concerns heat transfer around charged metallic nanostructures in electrolyte solutions. This situation is ubiquitous in biological environments, but also, in solar energy conversion devices where dissolved ions are present in order to control the stability and the state of dispersion of the nanoparticle suspension. However, the effect of these ions and their interaction with the surface charge on the nanoscale heat transfer has been only scarcely addressed[54]. In this manuscript, we report a two to threefold increase of G with the nanoparticle surface charge immersed in an electrolyte using molecular dynamics (MD) simulations. We investigate in detail the effects of ions and their concentration, nanoparticle wettability (hydrophilic and hydrophobic surface), and the surface charge and its distribution on the interfacial thermal conductance between the nanoparticle and its electrolyte solution. This allows us to describe heat transfer in terms of a master curve that depends only on the surface charge density. The role of water structuring and ion accumulation is here subsequently highlighted. The manuscript is structured as follows: In section II, we present the physical model considered in this work. In section III, we give computational details. Section IV is devoted to the presentation of the results and a discussion. The main results are summarized in the Conclusion.

## MATERIALS AND METHODS

The simulation domain is a cube having a side length of 7 nm, and the gold nanoparticle has a diameter of 3 nm, with a corresponding volume fraction ɸ=4%. Overall, the system contains 11400 water molecules and 887 gold atoms. Note that the gold atoms interact through non-bonded interactions. Initially, the water molecules are thermalized at 300 K and are randomly distributed around the spherical nanoparticle. Periodic boundary conditions are applied in all three spatial directions. A spring having a constant of $100 \frac{kcal}{mol}/Å^2$ is used to tether the nanoparticle's center of mass to the center of the simulation box. Therefore, the nanoparticle can rotate freely around its center while it can barely move in any direction. When considering salt in water, we use sodium chloride with a concentration of C=0.35 mol/L. Counter-ions are also added to the simulation box in addition to the salt ions to ensure the global electroneutrality of the whole system. Concerning the surface charge, six different cases are considered in our study:

I. Gold nanoparticle in pure Water (GW)
II. Gold nanoparticle (neutral surface) in Aqueous Electrolyte (GE)
III. Negatively Charged Gold nanoparticle (-1 e/nm$^2$ = -160 mC/m$^2$) in Aqueous Electrolyte (NG)
IV. Positively Charged Gold nanoparticle (+1 e/nm$^2$ = 160 mC/m$^2$) in Aqueous Electrolyte (PG)
V. Negatively Charged Gold nanoparticle (-2 e/nm$^2$ = -320 mC/m$^2$) in Aqueous Electrolyte (NNG)
VI. Positively Charged Gold nanoparticle (+2 e/nm$^2$ = 320 mC/m$^2$) in Aqueous Electrolyte (PPG)

The salt concentration and surface charges density are within a range of experimentally accessible values[47]. As mentioned before, hydrophilic and hydrophobic gold nanoparticles will be addressed in each case. Finally, heterogeneous (elementary charge on a fraction of randomly selected surface atoms) and homogeneous (equal charge on all surface atoms) distribution[48] were also considered. A sketch of the simulation domain for three selected cases is shown in Figure 1.

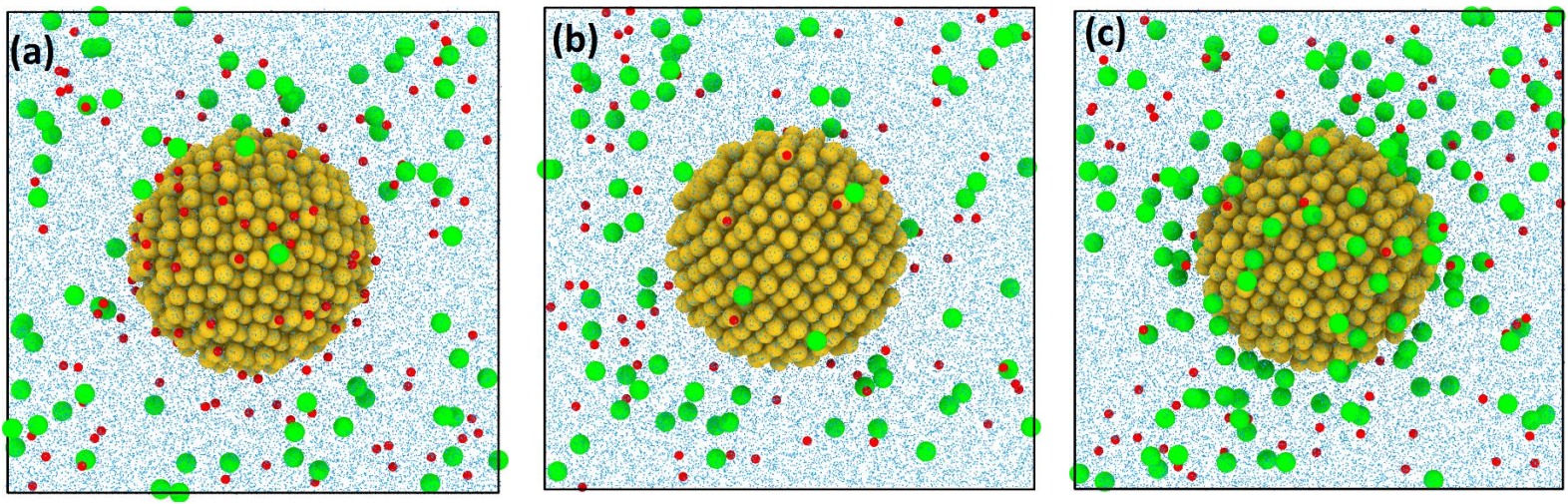

Figure 1. Sodium and Chloride ion distribution around homogeneously charged gold nanoparticles immersed in an electrolyte. Chloride and sodium atoms are shown in green and red respectively. (a) negatively charged nanoparticle (-320 mC/m$^2$), (b) uncharged nanoparticle, and, (c) positively charged nanoparticle (+320 mC/m$^2$) homogeneously charged nanoparticles. The difference in size between Sodium and Chloride ions is due to their different atomic radii.

We performed equilibrium molecular dynamics (EMD) simulations using the LAMMPS package[49]. The Open Visualization Tool (OVITO) is employed to visualize atomic configurations[50]. A Lennard-Jones (LJ) potential parametrization is considered to describe interatomic forces of gold atoms in the nanoparticle[51,52] while water is modeled using the TIP4P/2005 potential[53]. Ions were simulated with the scaled-ionic-charge Madrid model[54], using a scaling factor of 0.85. For consistency, the charge of nanoparticle atoms was rescaled with the same factor[55]. Although scaled-ionic models have been originally developed for bulk systems, they may be employed to model interfaces[55–59]. The non-bonded interactions between different atom types are also modeled using the LJ potential. Long-range Coulombic interactions were computed using the particle-particle particle-mesh (PPPM) method, and water molecules were held rigid using the SHAKE algorithm. More details about the interaction parameters and surface charges can be found in the supplementary material. Note that due to the presence of both co-ions and counter-ions, an electric double layer is generated around the charged nanoparticle[60]. This electric double layer is characterized by its Debye length equal to 0.5 nm for the salt concentration considered here. Such a value of the Debye length is smaller than the distance between the nanoparticle surface and the boundaries of the simulation box. The corresponding values of the surface potential may be estimated using Grahame's equation[60]:

$$V_S = \frac{k_B T}{e} \sinh^{-1}\left[\frac{\sigma}{120\ mC/m^2 \times \sqrt{cs/1\ mol/L}}\right] \quad (1)$$

where $k_B$ is Boltzmann's constant, T is the temperature, e is the elementary charge, σ the surface charge, and cs is the salt concentration. For the highest value of the surface charge, the surface potential varies between 85 mV and 200 mV on the range of values of salt concentration considered in this study. These values are realistic and importantly well below the thermodynamic window of water.

The Maxwell–Boltzmann velocity distribution is used to thermalize the atoms at the beginning of the simulation at 300 K. Then, a Nosé-Hoover thermostat is applied to the whole system for 1 ns to maintain the system at 300 K. After this equilibration period, the thermostat is removed and the simulation proceeds for 2 ns for the averaging process during which the density and temperature of each computational bin are calculated. Four distinct methods can be employed in the framework of MD simulations to compute the thermal interface conductance. Transient simulation with/without internal thermal resistance of the nanoobject[3,39], steady-state non-equilibrium simulation[61,62], and equilibrium molecular dynamics (EMD) simulation[34,35] have been extensively used in the literature. The accuracy of these methods was compared to the Kapitza conductance calculations of a nanoparticle immersed in water and it was shown that these four approaches yield close values of the thermal interface conductance[63]. The interfacial thermal conductance in EMD can be obtained based on the fluctuations of the temperature difference between the gold nanoparticle surface atoms and the surrounding aqueous solution (within the cut-off distance of the water and nanoparticle atomic interactions), using [64–66]:

$$G^{-1} = R_K = \frac{A}{k_B T_0^2}\int_0^\infty <\Delta T(t)\Delta T(0)>dt, \quad (2)$$

where $k_B$ is the Boltzmann constant, A is the nanoparticle surface area, $T_0$ is the equilibrium temperature of the system, and <> denotes an ensemble average. When the simulation starts, the density profile distribution of the nanoparticle shows that as the result of the interactions between gold atoms, the nanoparticle expands in such a way that its radius changes from 1.5 nm to 1.7 nm while keeping its spherical shape. Therefore, *A* is calculated based on this new value of the radius. Considering the first water shell around the nanoparticle which has a thickness of 1 nm, $\Delta T(t)$ is the instantaneous temperature difference between the nanoparticle and the first water shell. The convergence of the instantaneous temperature difference autocorrelation function for each case analyzed here is assessed on the fly as illustrated in the supplementary material in Figure S2. A timestep of 1 fs is used in all simulations (smaller timesteps were also tested, and yielded no significant difference). The calculated Kapitza conductance is averaged over 200 different initial independent configurations (obtained by considering different initial velocity distributions) for each different case shown here.

## RESULTS and DISCUSSION

As reference systems, we first consider uncharged gold nanoparticles immersed in water in the absence of ions. We have considered two different wetting situations, namely hydrophilic and hydrophobic nanoparticles. The interaction energies corresponding to the hydrophilic and hydrophobic cases are chosen such that the value of the gold/water contact angle is 30° and 90° respectively (see section III of the supplementary information for details on the contact angle measurements). Note that these values of the contact angle correspond to uncharged nanoparticles and that due to the presence of ions, the nanoparticle surface wettability may change. Regarding the Kapitza conductance, we have found 120±2 and 59±1 MW/m²/K for the hydrophilic and hydrophobic gold nanoparticles respectively. These values are in good agreement with experimental data obtained for a flat gold surface-water interface, which were 100±20 and 50±5 MW/m²/K for hydrophilic and hydrophobic surfaces respectively[67]. This observation validates our model system and methodology. The small deviations are somehow expected since the exact solid-liquid interactions for hydrophilic and hydrophobic surfaces are not specified in the experimental investigation[67].

Figure 2 shows the Kapitza conductance at the interface between charged gold nanoparticles and the aqueous electrolyte. For the uncharged system (GE), the Kapitza conductance is about 123 and 67 MW/m$^2$/K for hydrophilic and hydrophobic gold surfaces respectively. Note that, in the case of the uncharged nanoparticle, no electrostatic interaction takes place between the ions and the nanoparticle, and the interaction is limited to the Lennard-Jones contribution. Figure 3 shows that the water density distribution around the GW coincides with GE for hydrophobic surfaces and only a small difference is observed for the hydrophilic surface at a distance of 17.5 Å. This implies that the addition of the ions does not change the water density distribution significantly around uncharged nanoparticles. According to Figure 2, the small increase in the Kapitza conductance of GE compared to the corresponding GW case is related to the presence of the ions around the nanoparticle surface, which increases the overall interfacial conductance. Density layering occurs around the nanoparticle over a spatial extension limited to 8 Å from the nanoparticle surface. For larger distances, water density converges to its bulk value - 985±5 kg/m$^3$ - regardless of surface wettability and charge (see section II of the supplementary material for more details).

We now investigate the impact of the surface charge on interfacial heat transfer. Figure 2 shows that, as the absolute value of surface charge density ($\Sigma$) increases, the Kapitza conductance increases linearly. This increase is observed for both the hydrophilic and the hydrophobic nanoparticle. Compared to the interface between an uncharged nanoparticle and the aqueous electrolyte, it can be concluded that the hydrophilic surface with ±160 mC/m$^2$ surface charge density displays an enhancement of the Kapitza conductance by a factor of 1.5. Such an increase is a bit higher for the hydrophobic nanoparticles, falling in the range of 2~3. Figure 2C illustrates the relative increase of the Kapitza conductance with respect to the uncharged nanoparticle case. The data for which the increment ($\Delta G$) is negative correspond to an uncharged nanoparticle in pure water. It can be seen that the relative increase of the Kapitza conductance is independent of the surface wettability. Such observations enabled us to describe $\Delta G$ versus $\Sigma$ on a master curve defined by $\Delta G = \alpha \Sigma$ where $\alpha$ is expressed in units of (pW/eK). The parameter $\alpha$ for the negatively charged surface is $\alpha$ = -53 while for the positively charged surface, we have $\alpha$ = 48.15.

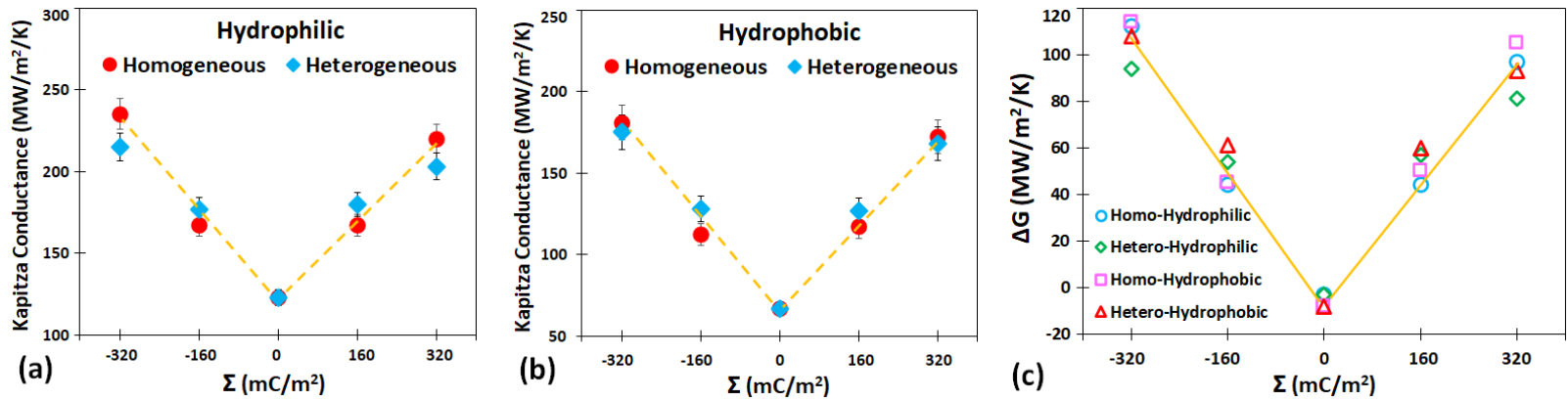

Figure 2. Kapitza conductance between the electrolyte and (a) hydrophilic and (b) hydrophobic gold surface for homogeneous and heterogenous charge distributions as a function of the surface charge density (Σ). (c) Difference between the Kapitza conductance of each case and corresponding hydrophilic or hydrophobic GE cases (ΔG), 123 and 67 MW/m²/K respectively, along with a master curve (yellow line) which predicts the general trend. "Homo" and "Hetero" denote homogenous or heterogenous charge distributions on the nanoparticle surface respectively.

Previous studies have discussed the role of liquid layering on the Kapitza conductance[45,68,69] at liquid-solid interfaces. In the following, we relate the enhancement of $G$ to the change of the water density distribution in the vicinity of the nanoparticle, which is induced by the surface charge. Figure 1a displays a snapshot of the sodium ions in the vicinity of the negatively charged nanoparticle, Σ=-320 mC/m². Compared to Figure 1b which shows a neutral surface, the accumulation of the sodium ions near the nanoparticle in Figure 1a can be explained by the electrostatic interaction between the surface charge and sodium ions. If we follow the motion of the sodium ions, we observe that, besides the accumulation near the nanoparticle surface, some ions are attached to the nanoparticle as a result of the strong electrostatic interaction. The accumulation of the chloride ions around the nanoparticle surface for Σ=+320 mC/m² can be seen in Figure 1c. However, no attachment of the chloride ions on the nanoparticle surfaces is observed. (See section II of the supplementary material for detailed information about the ions distribution profile).

The relative shift of the water distribution may be qualitatively understood by the dipolar interaction between the water molecules and the gold surface, as we explain in the following. Figure 3 shows that, for the homogeneously distributed charged hydrophilic gold nanoparticle, the water density profile is shifted towards the nanoparticle surface as compared to the uncharged nanoparticle, GE. This shift involves a layer of approximately 1-2 Å from the nanoparticle surface for hydrophilic and hydrophobic surfaces respectively. The density shift is more pronounced for large values of the absolute surface charge. The relative shift of the water distribution may be qualitatively understood by the following considerations. When the surface charge is positive, oxygen atoms are attracted by the nanoparticle due to both electrostatic and Lennard-Jones interactions, while the hydrogen atoms are repelled as a result of the net Coulombic interaction. However, the electric field created by the nanoparticle decreases in amplitude with the distance to its surface, and the attraction between the oxygen atoms and the surface dominates over the electrostatic repulsion exerted on the hydrogen atoms. Therefore, the attraction of water molecules toward the nanoparticle surface is driven by the Coulombic attraction between the oxygens and gold. Reciprocally, when the surface charge is negative, the Coulombic attraction between hydrogen and the surface compensates for the repulsion felt by the oxygens and explains the attraction of water molecules toward the charged surface.

Considering 17 Å for nanoparticle radius as discussed above, it is interesting also to notice that Figure 3 shows that regardless of surface charge, hydrogen and oxygen atoms penetrate in the outer layer of the nanoparticle surface. It can be observed that for the positively and negatively charged nanoparticles, the water density profiles start from 16.6 and 16.3 Å respectively, which means the water molecules get closer to the negatively charged nanoparticle surface. This can be attributed to the smaller Van der Waals atomic radii of hydrogen atoms compared to the oxygen atoms[70–72], which allows hydrogen atoms to get closer to negatively charged nanoparticles. Moreover, a bit higher value for the Kapitza conductance of negatively charged nanoparticle compared to the positively charged one observed in Figure 2c can be interpreted by the difference of atomic radii between H and O atoms on the one hand and between $Na^+$ and $Cl^-$ ions on the other hand. Positively charged ions have a smaller radius than negatively charged ions, therefore, they can come closer to the nanoparticle surface. At a constant value of the absolute surface charge, the Coulombic interaction between positive ions and the negatively charged nanoparticle is therefore stronger than the corresponding interaction between negative ions and the positively charged nanoparticle. This results in a decreased potential energy and a higher value of the Kapitza conductance.

Apart from the small difference in the start of the water density profile, it is interesting to notice that the water density shifts in the 1-2 Å thick layer from the nanoparticle surface do not depend on the sign of the surface charge. Note also that the second peak in the density profile has an amplitude that decreases when the surface charge increases in absolute value. As illustrated in Fig. S4 of the supplementary material, no meaningful relationship was found between the amplitude of this secondary peak and G. For this reason, we seek the following to relate the Kapitza conductance enhancement to the change of the water density profile occurring at short distances < 2 Å.

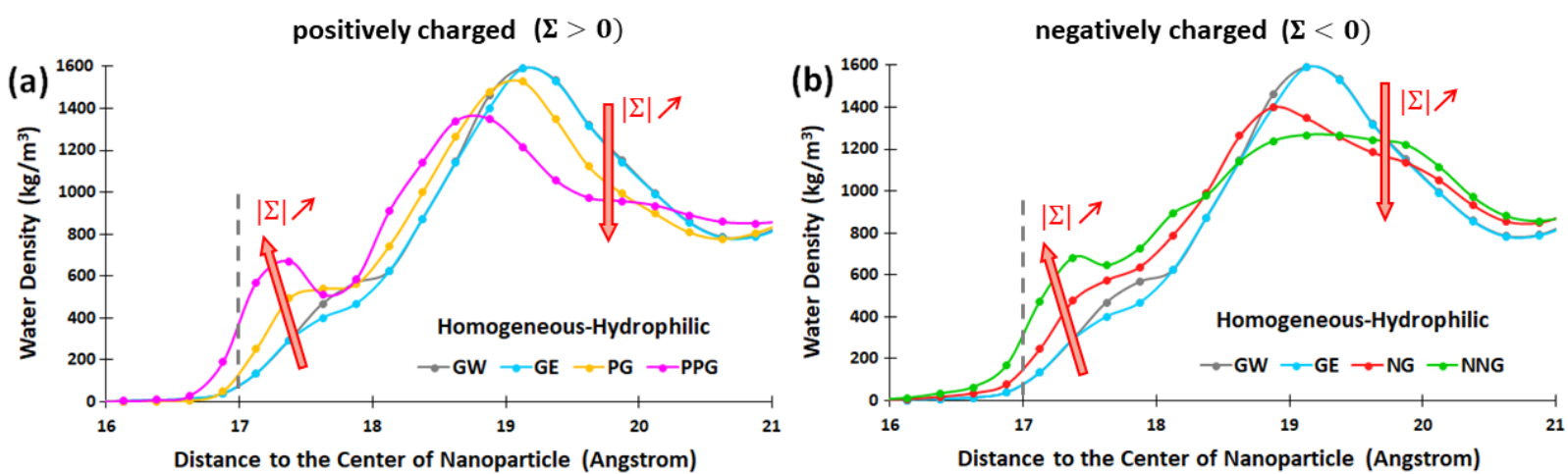

Figure 3. Water density distribution in the vicinity of the nanoparticle for which the nanoparticle surface is located at 17 Å and the nanoparticle center is located at the origin of the horizontal axis (at 0 Å). The dashed line denotes the nanoparticle surface. We considered gold nanoparticles in pure water (GW) and aqueous electrolyte as a function of the surface charge density (a) positive surface charge $\Sigma$ = +320 (PPG), +160 (PG), 0 (GE) mC/m$^2$ and (b) negative surface charge $\Sigma$ = -320 (NNG), -160 (NG), 0 (GE) mC/m$^2$.

In order to investigate this dependence in more detail, we define the accumulated density as the area between the water density curves and horizontal axis up to 17.5 Å, as illustrated in Figure 4a, approximately 1 Å from the nanoparticle surface. The accumulated density is a measure of the shift of the density profile towards the nanoparticle surface. Based on this idea, the accumulated normalized density is defined as follows: the accumulated density of neutral hydrophilic and hydrophobic uncharged nanoparticles (GE systems) is considered as the reference, as seen in Figure 4b. These two values are then subtracted from the accumulated density of the other hydrophilic and hydrophobic charged nanoparticles and the resulting value is divided by the bulk water density, i.e., 985 kg/m$^3$ (It can be observed in Figure 3 that the water density distribution of GW and GE are very close which implies that the reference cases shown in Figure 4 show GW cases too).

Figure 4b shows that one can identify different groups of systems depending on the value of the accumulated normalized density. The first group corresponds to uncharged nanoparticles, both hydrophilic and hydrophobic, and serves as reference cases. The second group is associated with an increment in ΔG of 40~60 MW/m$^2$/K which corresponds to |Σ|=160 mC/m$^2$ surface charges. The accumulated normalized density for this group varies between 0.5 and 1.5. The third group, for which the increment level in ΔG is 80~120 MW/m$^2$/K corresponds to |Σ|=320 mC/m$^2$. The accumulated normalized density for this group varies between 1 and 2. The overall dependence seen in Figure 4b suggests that the accumulated normalized density is one of the most determinant parameters driving the increment of ΔG.

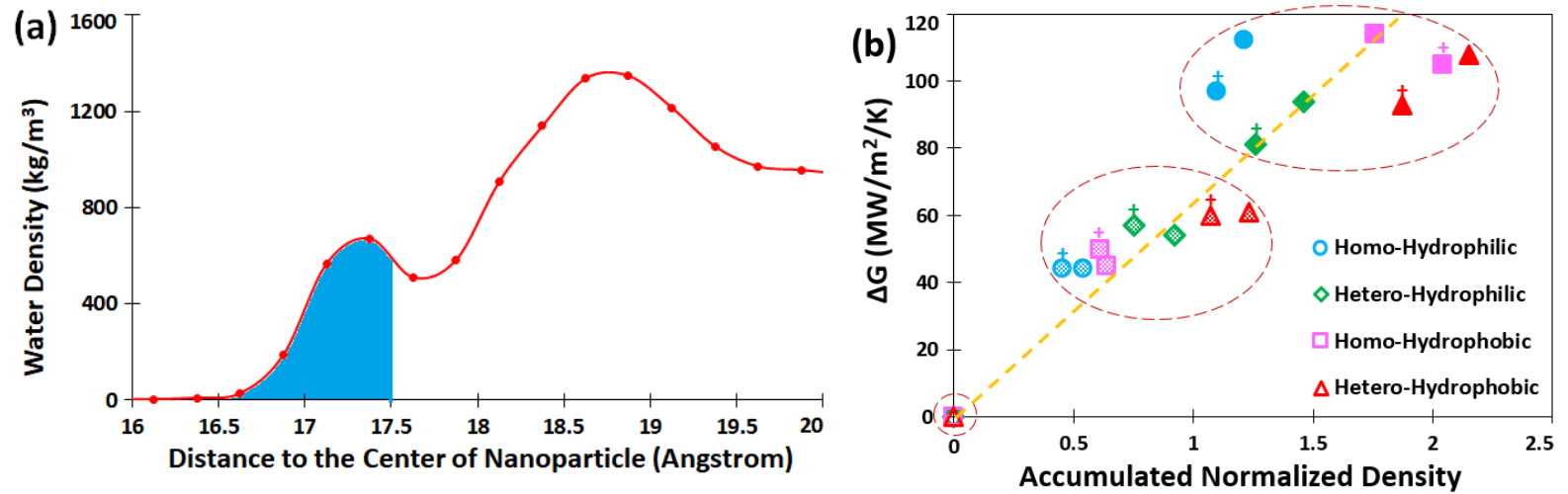

Figure 4. (a) Graphical illustration of the accumulated density and (b) Increment in Kapitza conductance (ΔG) with respect to the corresponding hydrophilic or hydrophobic GE cases (Σ=0), (123 and 67 MW/m$^2$/K) respectively, as a function of the accumulated normalized density of water. "Homo" and "Hetero" denote homogenous or heterogenous charge distributions on the nanoparticle surface respectively. Filled, patterned, and empty symbols denote |Σ|=320, 160, 0 mC/m$^2$ surface charges respectively. The "+" corresponds to a positively charged case while the other symbols refer to negatively charged systems.

However, it can be observed that some systems display similar accumulated normalized density but different values of ΔG. As an example, hydrophobic PG with heterogeneous charge distribution and hydrophilic PPG with homogeneous charge distribution are characterized by the same value of accumulated normalized density, around 1.08, while the Kapitza increment for the first is approximately half that of the second. Therefore, in addition to the accumulated normalized density, other microscopic parameters should play a role in the increase of thermal conductance. In order to identify these microscopic parameters, the contribution of the Lennard-Jones and Coulombic interaction to the total interfacial potential energy is shown in Figure 5, for the hydrophilic and hydrophobic gold nanoparticles. The potential energy is counted as positive in the case of attractive interactions and negative for repulsive interactions. Note that the integrated potential energy is higher in the hydrophilic case as compared to the hydrophobic system.

From the inspection of Figure 5, we can see that for uncharged nanoparticles, the potential energy is mainly controlled by the Lennard-Jones interactions between the oxygen atoms and the nanoparticle. The smaller proportion of the Lennard Jones interaction between the nanoparticle and the dissolved ions is due to their relative small number near the unharged nanoparticle. For charged nanoparticles, in addition to Lennard Jones interactions (shown in yellow), Coulombic interactions (in blue) contribute significantly to the interfacial potential energy. The Coulombic interactions include both water interactions with gold and ions interactions with gold. Coulombic interactions between water and a negatively charged nanoparticle is balanced between the attraction of the hydrogen atoms and the nanoparticle, and the repulsion of the oxygen atoms and the nanoparticle. Reciprocally, for a positively charged nanoparticle, Coulombic interactions between water and the nanoparticle are balanced between the attraction of the oxygen atoms and the nanoparticle, and the repulsion of the hydrogen atoms and the nanoparticle.

As it comes to the electrostatic interactions due to the ions, we can see from Fig. 5, that the corresponding values are not negligible, although the number of added ions is small as compared to the number of water molecules. This can be illustrated by the strong attraction between $Cl^-$ and $Na^+$ ions with the positively and negatively charged nanoparticles respectively. In fact, the Coulombic interaction energy between the counter-ions and the nanoparticle surface atoms has the same order of magnitude as the water/nanoparticle interaction energy. Note, however, that $Cl^-$ and $Na^+$ ions are repelled from the negatively and positively charged nanoparticle respectively and the corresponding Coulombic interaction is negligible as shown in Figure 5. A more detailed analysis of the interfacial potential energy can be found in section II.c of the supplementary material.

In summary, it is the strong Coulombic attraction between both the oxygen atoms and Cl$^-$ ions with a positively charged surface and the strong Coulombic attraction between both the hydrogen atoms and Na$^+$ ions with a negatively charged surface that is responsible for the enhancement of the interfacial bonding with respect to the uncharged nanoparticle. In other words, the potential energy between the nanoparticle and electrolyte is controlled by water-gold interactions (Lennard-Jones and Coulombic) and by the Coulombic interaction between the nanoparticle and the counter-ions.

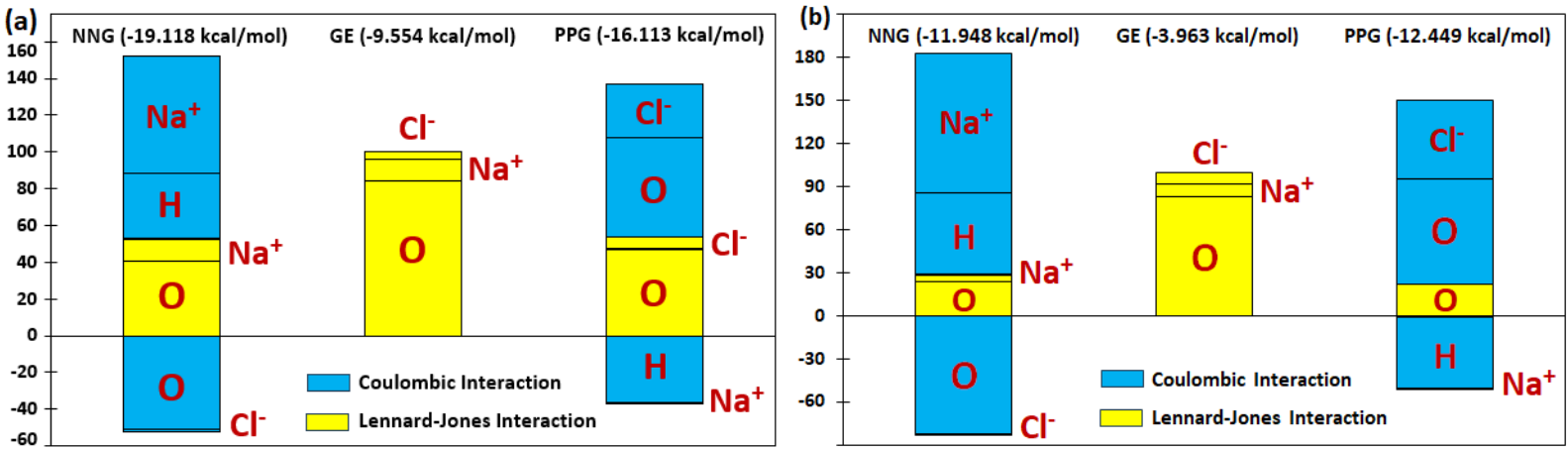

Figure 5. Decomposition (in percentage) of the total interfacial potential energy (given by the numbers in parentheses) for different surface charges densities (Σ=-320 (NNG), 0 (GE), +320 (PPG) mC/m$^2$) of hydrophilic (a) and hydrophobic (b) gold nanoparticle in C=0.35 mol/L electrolyte. Blue color corresponds to Coulombic interactions while yellow color corresponds to Lennard Jones interactions. O, H, Na$^+$, and Cl$^-$ denote the interfacial potential energy of oxygen. hydrogen, sodium, and chloride atoms with gold nanoparticle atoms respectively.

To confirm the role of the counter-ions, we investigate now the effect of the salt concentration on interfacial heat transfer. In the following, we consider a hydrophilic gold nanoparticle uniformly charged and three NaCl concentrations: C=0, 0.1, 0.35, and 1.0 mol/L. Note that, in the case C=0, there are only counter-ions in water. The variation of the Kapitza conductance as a function of the surface charge density is shown in Figure 6. It is interesting to note that, while the Kapitza conductance increases with the surface charge density, it is almost independent of the salt concentration.

Figure 5 clearly shows that a significant part of the interfacial energy potential originates from the contribution of H and O atoms. Since the number of water molecules is much larger as compared to the number of ions for all considered salt concentrations, it is expected that the contribution of H and O atoms to the total interaction energy should not change as the salt concentration varies. In terms of ions, the relative independence of G with the salt concentration is another evidence of the primary role of the counter-ions. We can propose the following interpretation to understand the effect of salt on heat transfer.

At a fixed value of the surface charge, when increasing the salt concentration, the number of ions increases, but only the counter-ions which compensate for the surface charge are involved in interfacial heat transfer. The minor role played by the co-ions has been already outlined in Figure 5: as an example, when C=0.35 mol/L, only $Na^+$ ions interact significantly with the negatively charged nanoparticle, while the $Cl^-$ co-ions contribute less to the potential energy as a result of their larger distance to the nanoparticle. The counter-ions in excess of the total nanoparticle surface charge are assumed not to participate in the interfacial potential energy and consequently in the heat transfer. This assumption is again consistent with the observed independence of heat transfer on salt concentration. These considerations are limited, however, to the maximal concentration C=1.0 mol/L considered here. For higher salt concentrations, the higher number of ions might affect the water molecules distribution near the nanoparticle surface and as a result the Kapitza conductance.

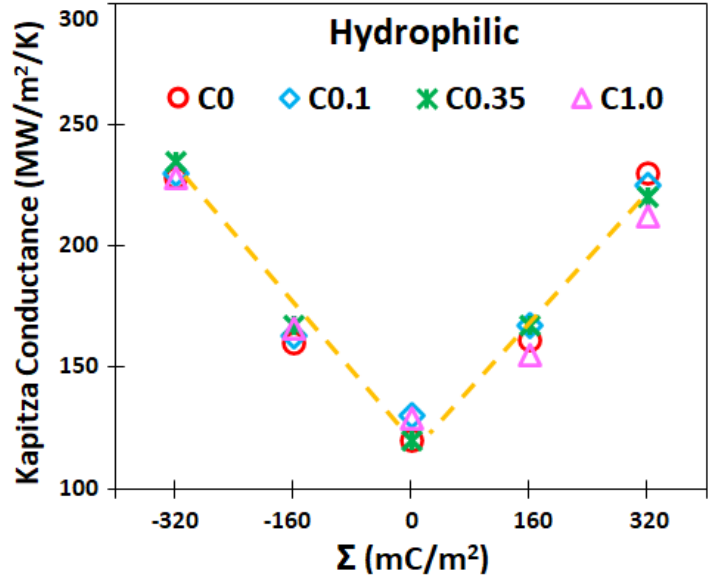

Figure 6. Variation of the Kapitza conductance as a function of the surface charge for different salt concentrations (C0, C0.1, C0.35, and C1.0 denote salt concentrations of 0, 0.1, 0.35, and 1.0 mol/L respectively). The nanoparticle considered here is hydrophilic.

## CONCLUSIONS

We studied the interfacial thermal conductance between a charged gold nanoparticle and a sodium chloride electrolyte solution using equilibrium molecular dynamics simulations. We investigated how interfacial heat transfer across electrically charged gold nanoparticles depends on the nanoparticle wetting properties and the spatial distribution of the surface charges. We showed that, for a neutral nanoparticle in an aqueous electrolyte, $G$ is approximately the same as if it were immersed in pure water. Considering uncharged hydrophilic/hydrophobic nanoparticles in the sodium-chloride solution as the reference system, we report a two to threefold increase of G for ±320 mC/m² surface charge density, for the hydrophilic and hydrophobic nanoparticles respectively. While the increment in $G$ for the charged surfaces is found to be almost independent of the surface wettability, the spatial charge distribution, and the salt concentration, the net surface charge density is found to be the primary parameter that affects the Kapitza conductance enhancement. This enables us to describe the evolution of the Kapitza conductance in terms of a master curve $\Delta G = \alpha |\Sigma|$ with $\alpha \approx 50$ pW/eK. We observed that the Kapitza conductance enhancement is not correlated to the magnitude of the first peak

of the water density profile. Rather, we report a correlation between G and the shift of the water density distribution toward the nanoparticle surface. Such a water density shift is a function of the surface charge. The shift of the water density profile and the accumulation of ions around charged nanoparticles strengthens the Coulombic interaction energy between the counter-ions and the nanoparticle, which consequently increases the Kapitza conductance. Besides, we have demonstrated that the Kapitza conductance is independent of the salt concentration, confirming thus the primary role played by the counter-ions. Overall, the large increase in *G* observed for charged nanoparticles could improve their performance for biomedical applications based on the use of plasmonic nanoparticles for which a reduction of treatment time is critical to confine heat to insane cells with the goal to minimize injury in healthy tissues. This study is also a first step to understanding the role of ions in heat transfer close to electrified surfaces.

## SUPPORTING INFORMATION

- Computational details
- Kapitza conductance calculations
- Distribution of water and ions around nanoparticle
- Interfacial potential energy
- Contact angle simulations

## ACKNOWLEDGMENTS

R.Rabani and M.H.Saidi wish to thank the financial support of the Iran National Science Foundation (INSF). S. Merabia acknowledges interesting discussions with D. Amans and Y. Yamaguchi. The simulations have been performed on the machines of the Pôle Scientifique de Modélisation Numérique (PSMN).

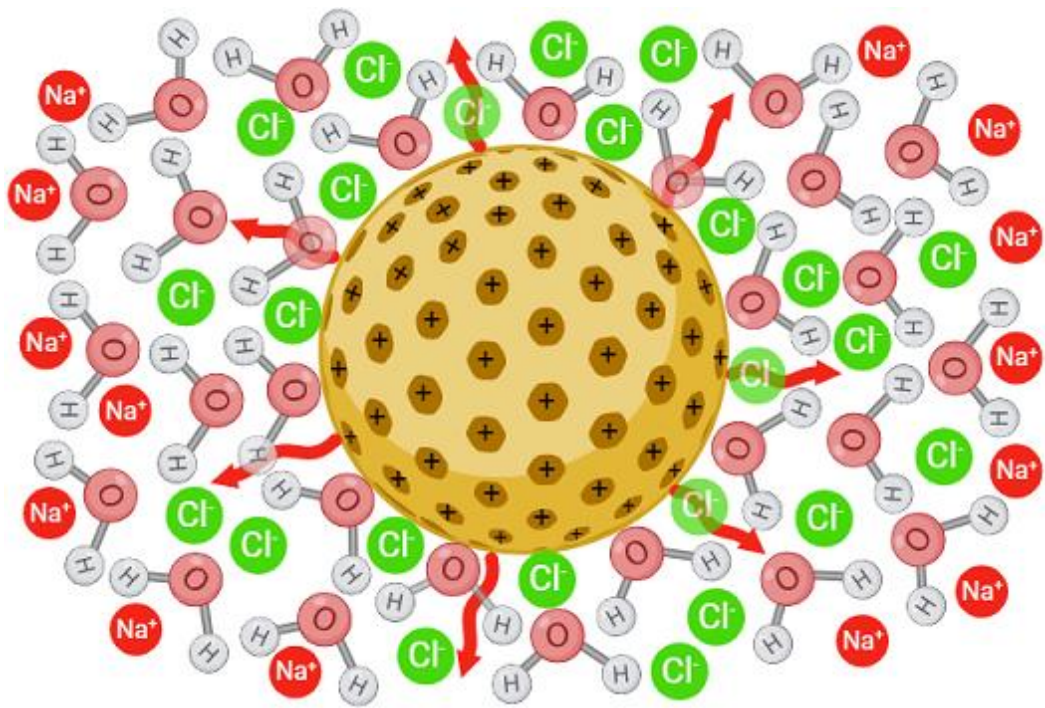

**TOC graphics**